\documentstyle[12pt]{article}
\begin{document}
\def\tj{\Theta_{1}}
\def\tjk{\Theta_{1}^{2}}
\def\td{\Theta_{2}}
\def\tdk{\Theta_{2}^{2}}
\def\pj{\partial_{1}}
\def\pd{\partial_{2}}
\def\dj{{\cal D}_{1}}
\def\dd{{\cal D}_{2}}
\def\djk{{\cal D}^{2}_{1}}
\def\ddk{{\cal D}^{2}_{1}}
\def\z#1{\zeta_{#1}}
\def\jmk{j_{-}^{2}}
\def\jpk{j_{+}^{2}}
\def\jmxx{j_{-xx}}
\def\jmx{j_{-x}}
\def\jpxx{j_{+xx}}
\newcommand\eq{(\ref{31})}
\def\be{\begin{equation}}
\def\ee{\end{equation}}

                \title{The Extended Supersymmetrization of the\\
                    Nonlinear  Schr\"{o}dinger  Equation}

                     \author{Ziemowit Popowicz
\\ \\
        Institute of Theoretical Physics, University of Wroclaw\\
        pl. M. Borna 9,  50-205 Wroclaw Poland \\
	e-mail ZIEMEK @ PLWRUW11.BITNET}
\date{}
\maketitle

\abstract{ We present the new extended supersymmetrization of the
          Nonlinear Schr\"{o}dinger Equation by introducing two superbosons
fields with the different gradation. Our model is different from this
considered by Toppan and in the reduced case (nonextended supersymmetry) is
also
different from this considered by Kulish and Roelofs and Kersten. We prove
that our model is integrable by presenting its Lax formulation.
}

\section{Introduction}
        The nonlinear solitonic equations have been widely studied since
the fundamental work of Gardner, Green, Kruskal and Miura [1-4] on  the
Korteweg-de Vries equation. It appeared that such equations describe a large
number of physical phenomenons and hence all possible generalizations
leading to soliton's behaviour were found to be are of great importance.
Interestingly the supersymmetrization gives us an alternative manner of
generalization [5].

The main idea of the supersymmetrization,  which comes from
quantum theory is to unify bosons and fermions. While connecting
supersymmetry with nonlinear equations,  one should distinguish between
two classes of supersymmetry - the so-called extended and nonextended
supersymmetry. In the nonextended case, one generalizes the given function
which describes the soliton behaviour by the addition of one (or several)
fermionic fields. As a result it was possible to generalize  [6-8]
many soliton equations  and obtain symmetries,  conservation laws,
prolongation structure, Lax pair and its bihamiltonian structure.
However in this generalization we encounter the problem of the interpretation
of new fermionic fields on the classical level. On the other hand this
effort is not an academic problem only.  The knowledge of the nonextended
supersymmetrization allows us,  in many cases to produce the extended
supersymmetrization.

We include in this case in addition to  fermions, also new bosons
fields.  After the supersymmetrization it appears that  the
so-called bosonic part of the model (in which all fermions vanish) give
us the new generalization of the model.  In such manner,  several soliton's
equation have been supersymmetrized, $KdV$ equation [8-11] and Boussinesq
equation [12-13] as examples.

In contrast,  the supersymmetrization of the Nonlinear Schr\"{o}dinger
Equation [7, 14-17]  has not drawn as much attention.  In fact, it was only
recently that the nonextended supersymmetrization of the Nonlinear
Schr\"{o}dinger Equation was proposed [7, 14-16]. In contrast to the
supersymmet
KdV equation it appeared that there exist only one supersymmetric extension of
this equation which is integrable.

Recently Toppan [17] has shown that the second hamiltonian structure
of the Nonlinear Schr\"{o}dinger Equation is connected with the reduced $sl(2)$
Kac-Moody algebra (in the Dirac sense).  This observation allowed him to obtain
the nonextended supersymmetrization of the Nonlinear Schr\"{o}dinger Equation.
However for the extended case (N=2) there are no supersymmetric extensions
of the Kac-Moody algebra [8] and therefore Toppan has proposed to use
two copies of chiral and antichiral (N=1) supercurrents in order to
construct the analog of the extended supersymmetric version of the Nonlinear
Schr\"{o}dinger Equation.

In this paper we present new extended supersymmetrization of the
Nonlinear Schr\"{o}dinger Equation.  Our idea is based on the introduction of
tw
supermultiplets with the different gradation.  Namely, we introduce two
supercurrents of the gradation one and zero respectively.  Only in the
special bosonic limit (described below) we recover the usual Nonlinear
Schr\"{o}dinger Equation.  Our model is different then this considered by
Toppan
Moreover after reducing our model to the nonextended case we obtain the
model which is different from the one considered by Kulish [7],  Roelofs and
Kersten [14] and by Brunelli and Das [16].  Our model in this nonextended case
corresponds to the missing possibilities in the Roelofs and Kersten
discussion - to the mixing gradation  of the fermion fields.

All results presented in this paper were obtained due to the extended
application of the symbolic computer language REDUCE - namely the Lax pair
for our model.  The existence of such pair  proves that our model is integrable
and yields the presentation of conservation laws together with its first
supersymmetric hamiltonian structure.

\section{Supersymmetric Nonlinear Schr\"{o}dinger
\protect\linebreak Equation}

        The basic object in the supersymmetric analysis is the superfield
and the supersymmetric derivative.  The superfields are the superfermions or
the superbosons depending,  in addition to
$x$    and $t$,  upon two anticommuting
variables,  $\tj$    and $\td$
$(\td \tj  =  - \tj \td , \  \tjk = \tdk = 0 )$.  Its Taylor
expansion with respect to  $\Theta$`s is
\be
\Phi \left( x, \tj , \td \right) = \omega(x)
+ \tj \z1 (x) + \td \z2 (x) + \td \tj u(x)\ ,  \label{1}
\ee
where the fields   $w,u$   are to be interpreted as the bosons (fermions)
fields while   $\z1 , \z2$    as the fermions (bosons) for the superbosons
(superfermions) field    $\Phi$.  The superderivatives are defined as
\be
\dj = \partial_{\tj} + \tj \partial_{x} \ ,
\qquad \dd = \partial_{\td} + \td \partial_{x} \ ,
\label{2}
\ee
with the properties
\be
\dd \dj + \dj \dd = 0 \ ,
\label{3}
\ee
\be
\djk = \ddk = \partial_{x} \ .
\label{4}
\ee

Below we shall use the following notation
$\left( {\cal D}_{i} \phi\right)$     denotes the
outcome of the action of the superderivative on the superfield,  while
${\cal D}_{i} \Phi$
        denotes the action itself of the superderivative on the superfield.
        We use in our construction two superboson fields

\be
F = j_{+} + \tj \z1 + \td \z2 + \td \tj f \ ,
\label{5}
\ee
\be
G = g + \tj \eta_{1} + \td \eta_{2} + \td \tj j_{-} \ .
\label{6}
\ee
Our construction heavily relies on the gradation of (6) and
(7).  We choose
zero as the gradation for $G$ while the one for $F$ with the following
prescription
\begin{eqnarray}
       \deg(x)= -1  \ ,  &   \deg(t)=-2 \    ,
\qquad &  \deg(\Theta)=-1/2 \ ,
\label{7}
\\
\nonumber \\
  \deg(j_{+}   )=1\ ,  &
     \deg(\zeta  )=3/2 \   ,
& \deg(f   )=2\  ,
\label{8}
\\
\nonumber \\
\deg(j_{-}   )=1\ ,
& \deg(\eta  )=1/2 \   ,  \qquad
&\deg(g  )=0 \ .
\label{9}
\end{eqnarray}

The extended supersymmetric Nonlinear Schr\"{o}dinger Equation which we would
like to study has the form
\be
\dot{F} = F_{xx} - G^{2} F^{3}
+ 2F \left( \dj \dd \left( GF\right)\right) \ ,
\label{10}
\ee
\be
   \dot{G} = - G_{xx}
+ G^{3}F^{2}
- 2G \left( \dj\dd \left( GF\right)\right) \ ,
\label{11}
\ee
where the dot denotes the time derivative while $x$ the space derivative.

Notice that the superboson $G$ has different gradation than
$F$,  hence
our equation (10-11) could not be written as a single equation.  In other
words the superboson $G$ is not complex conjugate to  superboson $F$.
On the other
hand  these equations  can be written  in the component form using (5-6),
but the formulas are complicated.  We can investigate the properties of
our generalization and we present two particular reductions,   from which we
obtain the usual Schr\"{o}dinger equation:

{\bf a.} \ The bosonic limit in which all fermions vanish
\begin{eqnarray}
&\dot{j_{+}}
& = \jpxx - j_{+}^{3} g^{2}
- 2j_{+}gf + 2 \jpk j_{-} \ ,
\label{12}
\\
\nonumber \\
& \dot{f}
& =   f_{xx} + 2gf^{2} - 2\jpxx j_{-} g -
4 j_{+x} j_{+} g_{x}
\nonumber \\
 \\
& &
\quad - 2\jpk g_{xx} - 3 \jpk g^{2}  f - 2j_{-} j^{3}_{+}g
 +   2 j_{-} j_{+} f \ ,
\nonumber \\
\nonumber \\
\label{13}
& \dot{g}
&  =
 - g_{xx} - 2 g^{2}f + \jpk g^{3} - 2 j_{-} j_{+} g \ ,
\label{14}
\\
\nonumber \\
&\dot{j_{-}}
&  =
 - \jmxx
+ 2 \jmxx g^{2} + 4 j_{-x} g_{x} g + 2 j_{-} g_{xx} g +
\nonumber \\
 \\
& &  \quad 2 j_{-} g^{3} f - 2 \jmk j_{+} - 2 j_{-}
gf + 3 j_{-} \jpk g^{2} \ .
\nonumber
\end{eqnarray}
%%\label{15}
%\ee
Moreover these equations could be reduced further by demanding that
 $f  = 0$  and
$g =0$ or
$j_{+}    =0$  and
$j_{-}     =0$.  In the first case we obtain
the usual Nonlinear Schr\"{o}dinger Equation
\be
\dot{j_{+}} = \jpxx + 2 \jpk j_{-} \ ,
 \label{16}
\ee
\be
\dot{j_{-}} = - \jmxx - 2 \jmk j_{+} \ ,
 \label{17}
\ee
assuming that
$j^{*}_{-} = j_{+}$
          where
$*$     denotes the complex conjugation.
In the second case we obtain the following equations
\be
\dot{f} = f_{xx} + 2 g f^{2} \ ,
\label{18}
\ee
\be
\dot{g} = - g_{xx} - 2 g^{2} f\ ,
\label{19}
\ee
which are similar to the equations (16) and (17).  However these could
not be interpreted as the Nonlinear Schr\"{o}dinger Equation because
$f$  is not complex conjugate to
$g$   due to the different gradation.

{\bf b.}\  Nonextended case is obtained assuming that
\be
F = F_{1} = j_{+} + \Theta_{1} \zeta \ ,
\label{20}
\ee
\be
   G = \Theta_{2} G_{1}
 = \Theta_{2} \left(
\eta + \Theta_{1} j_{-} \right) \ ,
  \label{21}
\ee
which reduces our equations (10-11) to
\be
\dot{F}_{1} = F_{1xx} + 2 F_{1} \left(
\dj \left( G_{1} F_{1} \right)\right) \ ,
\label{22}
\ee
\be
   \dot{G}_{1} = - G_{1xx}
- 2 G_{1} \left( \dj \left( G_{1}
F_{1} \right) \right) \ .
  \label{23}
\ee
In the component these equations take the form
\begin{eqnarray}
&\dot{j_{+}} & = \jpxx + 2 \jpk j_{-} + 2 j_{+} \zeta
\eta \ ,
\label{24}
\\
\nonumber \\
&\dot{\zeta} & = \zeta_{xx} + 2 j_{+}
\left( j_{+} \eta \right)_{x} + 2 \zeta j_{+} j_{-} \ ,
\label{25}
\\
\nonumber \\
&\dot{\eta} & = - \eta_{xx} - 2 \eta j_{+} j_{-} \ ,
\label{26}
\\
\nonumber \\
& \dot{j_{-}} & = - j_{xx}
- 2 \jmk j_{+} - 2 \zeta \eta j_{-}
- 2 \eta \eta_{x} j_{x} \ ,
\label{27}
\end{eqnarray}
from which it immediately follows  that our equations are different from
these considered by Kulish [7],  Roelofs and Kersten [14].

\section{The Lax and Hamiltonian Formulations
\protect\linebreak
and Conservation Laws}

	The Nonlinear Schr\"{o}dinger Equation has been solved via the inverse
scattering transformation by Zakharov and Shabat [18].  Later,  it appeared
that
this equation could be solved in the scheme of the AKNS method [4].  Quite
recently it appeared that this equation could be formulated in  terms of the
scalar as well as matrix Lax pair [16-17].  The Nonlinear Schr\"{o}dinger
Equation is the bihamiltonian equation e.q. it can be written down as
\be
{d \over dt } {j_{+}\choose j_{-} }
= P_{2} \cdot \mbox{\em grd}\,  H_{2} = P_{1} \cdot \mbox{\em grd}\,
H_{3} \ ,
\label{28}
\ee
where ${\mbox{\em grd}}$ denotes the gradient of the functional
\begin{eqnarray}
&H_{3} & = \int dx\, \left( J_{+ x} J_{- x} +
J_{+}^{2}  J_{-}^{2} \right) \ ,
\label{29}
\\
\nonumber \\
&   H_{2}& = \int dx\, J_{+ x} J_{-} \ ,
\label{30}
\end{eqnarray}
while the $P$ is the Poisson tensor
\be
P_{1}
=       \pmatrix{ 0 & 1 \cr
-1 & 0 } \ ,
\label{31}
\ee
\be
P_{2} =
\pmatrix{ -2J_{+}\partial^{-1} J_{+} \ ,
&\partial_{x} + 2 J_{+} \partial^{-1} J_{-} \cr\cr
\partial_{x} + 2J_{-} \partial^{-1}J_{+} \ ,
& - 2 J_{-} \partial^{-1} J_{-} }
\ .
\label{32}
\ee

The $P_{2}$   operator could be obtained from the Kac-Moody $sl(2)$
algebra,  if we use the Dirac reduction procedure.  Indeed let $P$
\be
P =
\pmatrix{ 0 & \partial_{x} - 2 J_{0} & -J_{+}
\cr\cr
\partial_{x} + 2 J_{0} & 0 & J_{-} \cr\cr
J_{+} & - J_{-} & - {1\over2} \partial_{x}  }
\ ,
\label{33}
\ee
be a Poisson tensor which corresponds to the Kac-Moody algebra. Now let us
briefly explain the standard  Dirac reduction  [19]  formula.  Let $U,V$  be
two linear spaces with coordinates $u$ and $v$. Let
\be
P(u,v) = \left[
\begin{array}{cc}
P_{uu}\ , & P_{uv} \cr\cr
P_{vu} \ , &P_{vv}
\end{array}
\right]
\ ,
\label{34}
\ee
be a Poisson tensor on   $U\oplus V$.  Assume that
$P_{vv}$           is
invertible,  then
\be
P = P_{uu} - P_{uv} P_{vv}^{-1}
P_{vu} \ ,
\label{35}
\ee
is a Poisson tensor on $U$.  Now it is easy to confirm yourself that the
application of the formula (35) to the $sl(2)$ Kac-Moody algebra yields
the formula (32).

	In the next we will consider the scalar Lax pair of the Nonlinear
Schr\"{o}dinger Equation.  To state the problem, let us first briefly describe
the KP hierarchy following [2] and the convention introduced there.
The pseudodifferential Lax operator is defined by
\be
L = \partial_{x} + \sum\limits^{\infty}_{i=0}
U_{i} \partial^{-i} \ ,
\label{36}
\ee
where the
$u_{i}$    are an infinite set of fields depending on the spatial
coordinate
$x$
and the time parameters    $t_{k}$.
The infinite set of differential
equations for the fields
$U_{i}$     is introduced via the equations
\be
{\partial L \over \partial t_{k}}
=  \left[ L^{k}_{+} , L\right] \ ,
\label{37}
\ee
where $+$    denotes the purely differential part of the k-th power of the
Lax operator.  The quantities
\be
F_{k} = {1 \over k}
Res\,  L^{k}
\label{38}
\ee
are first integrals of motion for the flows (37) . Here res denotes the
integral
$Res\, A = \int dx\, a_{-1}$
               for the generic pseudodifferential operator
$A = \ldots + a_{-1} \partial^{-1} + \ldots \ $.

For the Nonlinear Schr\"{o}dinger Equation we choose the scalar Lax pair
[17] as
\be
L = \partial_{x} + J_{-} \partial^{-1} J_{+} \ ,
\label{39}
\ee
and  our equation could be written down as
\be
   \dot{L} = \left[ L^{2}_{+} , L\right] \ .
     \label{40}
\ee

For the extended (N=2) supersymmetric case we consider the
super pseudodifferential Lax operator [9,10] in the form (36) where now
\be
U_{i} = b1_{n} + f1_{n}
\dj + f 2_{n}\dd + b2_{n} \dj \dd \ ,
\label{41}
\ee
and the $b1$
    and $b2$
		    are the superbosons while
				$f1$
				   and
					 $f2$
					     are superfermions.
	In order to find the supersymmetric Lax pair which gives us
the Superextension of the Nonlinear Schr\"{o}dinger Equation we can postulate
the most general form of this pair and try to fix it by computing the
equality (40).  We used the symbolic manipulation program REDUCE and
found that the following Lax operator
\be
L = \partial_{x} + G \cdot \partial^{-1} \cdot \dj \dd F \ ,
\label{42}
\ee
gives us the equations (10-11).

	This Lax operator could be reduced to the bosonic case or to the
nonextended case.  For the first case we obtained
\be
L = \pmatrix{ \partial_{x} + g \partial^{-1} f \ ,
& g\partial^{-1} j_{+} \cr\cr
- g \cdot \partial_{x} j_{+} + j_{-} \partial^{-1}f \ ,
&\partial_{x} + j_{-} \partial^{-1} j_{+}
} \ ,
\label{43}
\ee
which gives us the equations (12-15). For the second case we have
\be
L = \pmatrix{\partial_{x} - \eta \partial^{-1} \zeta \ ,
& \eta \partial^{-1} j_{+} \cr\cr
-j_{-} \partial^{-1} \zeta \ ,
& \partial_{x} + j_{-} \partial^{-1} j_{+} }
\ ,
\label{44}
\ee
with the equations (24-27).
	First, let us present four nontrivial conservation laws for
our model.   In order to do it we modify the formula (38) for which
\be
A = \ldots + b2_{-1} \dj \dd \partial^{-1} + \ldots
\label{45}
\ee
and therefore       $a_{-1} = b2_{-1}$,
\begin{eqnarray}
&H_{1}& = \int dX \, G\cdot F \ ,
\label{46}
\\
\nonumber \\
&H_{2}& = \int dX \, G_{x} F \ ,
\label{47}
\
\nonumber \\
&H_{3}& = \int dX\, \left\{ G_{xx} F - {1\over 3}
G^{3}F^{3} +
GF\left( \dj \dd \left( GF \right)\right) \right\} \ ,
\label{48}
\\
\nonumber \\
&H_{4}& = \displaystyle\int dX \, \bigg\{ G_{xxx} F +
{1 \over 2} G^{2} F^{2} \left( GF_{x} - G_{x} F\right) \ +
\cr\cr
&& \quad \displaystyle{3\over 2} \left( G_{x}F
- GF_{x}\right)
\left( \dj \dd \left( G \cdot F \right)\right) \bigg\} \ ,
\label{49}
\end{eqnarray}
where now $dX = dx d\Theta_{2}d\Theta_{1}$
                and our integral is understand as the Berezin
integral.

	Finally let us discuss the hamiltonian formulation of our
supersymmetric extension.  Our supersymmetric extension can be
written down as the
\be
{d \over dt} {F \choose G} = P_{1}\cdot \mbox{\em grd}\, H_{3} \ ,
\label{50}
\ee
where,  interestingly the Poisson tensor has the same form as in (31).
This formulation gives us the first hamiltonian structure.  Unfortunately
we could not find its second structure,  what is probably connected with
the nonexistence of the N=2 supersymmetric $sl(2)$ Kac-Moody algebra.

\end{document}